\documentclass[twocolumn,showpacs,prl,superscriptaddress]{revtex4-1}%
\usepackage{amsfonts}
\usepackage{amsmath}
\usepackage{amssymb}
\usepackage{graphicx}
\usepackage{txfonts}
\usepackage{xcolor}%
\providecommand{\U}[1]{\protect\rule{.1in}{.1in}}

\begin{document}
\title{Characterization of arbitrary-order correlations in quantum baths by weak measurement}
\author{Ping Wang}
\affiliation{Department of Physics, The Chinese University of Hong Kong, Shatin, New Territories, Hong Kong, China}
\author{Chong Chen}
\affiliation{Department of Physics, The Chinese University of Hong Kong, Shatin, New Territories, Hong Kong, China}
\author{Xinhua Peng}
\affiliation{Hefei National Laboratory for Physical Sciences at Microscale, Department of Modern Physics,
and Synergetic Innovation Center of Quantum Information \& Quantum Physics,
University of Science and Technology of China, Hefei 230026, China}
\author{J\"{o}rg Wrachtrup}
\affiliation{3rd Institute of Physics, Research Center SCoPE and IQST, University of Stuttgart, 70569 Stuttgart, Germany}
\affiliation{Max Planck Institute for Solid State Research, Stuttgart 70569, Germany}
\author{Ren-Bao Liu}
\email{Corresponding author. Email: rbliu@cuhk.edu.hk}
\affiliation{Department of Physics, The Chinese University of Hong Kong, Shatin, New Territories, Hong Kong, China}
\affiliation{Centre for Quantum Coherence, The Chinese University of Hong Kong, Shatin, New Territories, Hong Kong, China}

\date{\today}

\begin{abstract}
Correlations of fluctuations are the driving forces behind the dynamics and thermodynamics in quantum many-body systems.
For qubits embedded in a quantum bath, the correlations in the bath are the key to understanding
and combating decoherence - a critical issue in quantum information technology. However, there is no systematic method for
characterizing the many-body correlations in quantum baths beyond the second order or the Gaussian approximation.
Here we present a scheme to characterize the correlations in a quantum bath to arbitrary order. The scheme employs weak measurement of the bath via projective measurement of a central system.
The bath correlations, including both the ``classical'' and the ``quantum'' parts, can be reconstructed from the correlations of the measurement outputs.
The possibility of full characterization of many-body correlations in a quantum bath forms the basis for optimizing quantum control against decoherence in realistic environments, for studying the quantum characteristics of baths,
and for quantum sensing of correlated clusters in quantum baths.
\end{abstract}

\maketitle

\textit{Introduction\textemdash{}} The correlations of fluctuations
are the driving forces underlying the quantum dynamics and thermodynamic
processes (such as critical phenomena) of quantum many-body systems.
Conventionally the correlations in many-body physics are considered
at the second order or in the Gaussian approximation (which amounts
to taking into consideration the quasi-particle excitations around
the mean field), with the assumption that the higher order correlations are usually much smaller than the second order ones. Recent studies have revealed the importance of higher
order correlations, especially in mesoscopic quantum systems~\cite{AlvarezScience2015,SchweiglerNature2017}.
The study of higher-order correlations, however, is challenging due
to their many-body nature.

The quantum many-body correlations are particularly important to quantum information technology
for their relevance in decohernece of central quantum systems embedded
in quantum baths~\cite{WitzelPRB2006,CywinskiPRL2009,YangPRB2008a,YaoPRB2006,YangRPP2017}.
Recently, Gasbarri and Ferialdi~\cite{GasbarriPRA2018} show that the
dynamics of a central quantum system is determined by the correlations
in the quantum bath and the effects of a quantum bath can
be fully simulated by \textit{complex} classical noises. This remarkable
work paves the way of optimal quantum control for quantum gates and
quantum memory in realistic environments. For classical noises, once
the noise spectra are known, the quantum control of the central quantum
systems can be designed to combat the decoherence~\cite{KofmanPRL2001,KofmanPRL2004,CywinskiPRB2008}.
Outstanding examples are dynamical decoupling~\cite{Viola1998,Ban1998,Zanardi1999,ViolaPRL1999,KhodjastehPRL2005}
and dynamically optimized quantum
gates~\cite{GraceJPB2007,GordonPRL2008,WestPRL2010,LiuNatCommun2013}.
For quantum baths, the back-action of the central system means that the bath correlations may need to be characterized each
time for each new quantum operation and yet the optimization would
be extremely time-consuming, if not impossible at all, due to the
notoriously difficult quantum many-body problems. Now thanks to the
progress in Ref.~\cite{GasbarriPRA2018}, the quantum control optimization
can be applied to quantum baths as well, as long as the bath correlations
can be characterized.

Thus, both for studying many-body physics and for applications in quantum information technology,
highly desirable is a systematic method to measure the many-body correlations in a quantum bath.
With the assumption of Gaussian noises, the noise
spectroscopy (e.g., by dynamical decoupling or frequency combing)~\cite{AlvarezPRL2011,YoungPRA2012,PazPRL2014,NorrisPRL2016,PazPRA2017,FerrieNJP2018}
can be employed to obtain the noise correlation spectra. The applications
of the frequency comb approach, e.g., to higher-order correlations,
however, are tricky due to the interference of nonlinear effects~\cite{AlvarezPRL2011,PazPRL2014,NorrisPRL2016}  and
 spurious signals~\cite{LoretzPRX2015}
and yet are often limited to the case of pure dephasing~\cite{NorrisPRL2016}.

In this paper, we present a general scheme for completely characterizing
the correlations in a quantum bath. The scheme is based on weak measurement of the bath via projective measurement
of the central system. By designing the measurement sequence, the bath
correlations at arbitrary orders can be reconstructed from the correlations
of the measurement outputs. Quantum weak measurement has been used
to monitor quantum coherent oscillations~\cite{Korotkov2001,KorotkovPRB2001a},
characterize spectral diffusion~\cite{SallenNP2010}, and measure the
non-sysmmetric correlations~\cite{BednorzPRL2013,BueltePRL2018}. Multitime correlations of continuous weak measurements have also been studied~\cite{HaegelePRB2018,AtalayaPRA2018,TilloyPRA2018}.  Recently,
weak measurement was considered for improving spectral resolution in quantum
sensing~\cite{Retzker2018PRA,PfenderNC2019}. The application of weak measurement enabled
the high-spectral-resolution magnetic resonance spectroscopy of single nuclear
spins~\cite{PfenderNC2019}, which is possible due to the fact
that the disturbance to the
system caused by the weak measurement  (i.e., measurement induced decoherence)  is negligible. The weak disturbance
feature of weak measurement is exploited in our scheme of characterizing
bath correlations.

Before proceeding to present our scheme, here we first summarize
the main results of Ref.~\cite{GasbarriPRA2018}. A general Hamiltonian
$H=H_{0}+V$ is considered, where $H_{0}=H_{{\rm S}}(t)+H_{{\rm B}}$ contains the
system Hamiltonian $H_{{\rm S}}(t)$ (which may be time
dependent due to external control) and the bath Hamiltonian $H_{{\rm B}}$,
and $V=\sum_{\alpha}S_{\alpha}B_{\alpha}$ is the coupling between the bath operators $B_{\alpha}$ (the noise fields)
and the system operators $S_{\alpha}$. In the interaction picture,
\[
\hat{V}(t)=\sum_{\alpha}\hat{S}_{\alpha}(t)\hat{B}_{\alpha}(t),
\]
where the operator in the interaction picture is given by $\hat{A}(t)\equiv U_{0}^{\dagger}(t)AU_{0}(t)$
with $U_{0}(t)\equiv\mathcal{T}e^{-i\int_{0}^{t}H_{0}(\tau)d\tau}$
and $\mathcal{T}$ denoting time-ordering. The initial state of the
system and the bath is assumed to be separable, described by the density
operator $\rho(0)=\rho^{{\rm S}}(0)\otimes\rho^{{\rm B}}$. The density
operator in the interaction picture, $\hat{\rho}(t)\equiv U_{0}^{\dagger}(t)\rho(t)U_{0}(t)$,
evolves according to $\hat{\rho}(t)=\mathcal{T}e^{\int_{t_0}^{t}{\mathcal{L}}(\tau)d\tau}\hat{\rho}(t_0)$, with the Liouville superoperator $\mathcal{L}$
defined by $\mathcal{L}(t)\hat{\rho}=-i\left[\hat{V}(t)\hat{\rho}-\hat{\rho}\hat{V}(t)\right]$.
Defining the superoperators $\mathcal{A}^{\pm}$ as $\mathcal{A}^{-}\hat{B}=-i\left(\hat{A}\hat{B}-\hat{B}\hat{A}\right)/2$ (essentially a commutator)
and $\mathcal{A}^{+}\hat{B}=\left(\hat{A}\hat{B}+\hat{B}\hat{A}\right)/2$ (essentially an anti-commutator)
and using the identity $-i\left[\hat{A}\hat{B},\hat{C}\right]=2\left(\mathcal{A}^{+}\mathcal{B}^{-}+\mathcal{A}^{-}\mathcal{B}^{+}\right)\hat{C}$,
one obtains the reduced density operator of the central system $\hat{\rho}^{{\rm S}}(t)\equiv{\rm Tr}_{{\rm B}}\hat{\rho}(t)$
as
\begin{equation}
\begin{gathered}\hat{\rho}^{{\rm S}}(t)=\sum_{N=0}^{+\infty}\frac{2^{N}}{N!}\sum_{\{\alpha_{i}\},\{\eta_{i}=\pm\}}\int_{0}^{t}dt_{1}dt_{2}\cdots dt_{N}C_{\alpha_{N},\ldots,\alpha_{2},\alpha_{1}}^{\eta_{N},\ldots,\eta_{2},\eta_{1}}\\
\times\left[\mathcal{T}\mathcal{S}_{\alpha_{N}}^{\bar{\eta}_{N}}(t_{N})\cdots\mathcal{S}_{\alpha_{2}}^{\bar{\eta}_{2}}(t_{2})\mathcal{S}_{\alpha_{1}}^{\bar{\eta}_{1}}(t_{1})\right]\hat{\rho}^{{\rm S}}(0),
\end{gathered}
\end{equation}
determined by the bath field correlations
\begin{equation}
C_{\alpha_{N},\ldots,\alpha_{2},\alpha_{1}}^{\eta_{N},\ldots,\eta_{2},\eta_{1}}={\rm Tr}\left[\mathcal{T}\mathcal{B}_{\alpha_{N}}^{\eta_{N}}(t_{N})\cdots\mathcal{B}_{\alpha_{2}}^{\eta_{2}}(t_{2})\mathcal{B}_{\alpha_{1}}^{\eta_{1}}(t_{1})\rho^{{\rm B}}\right],
\end{equation}
where $\bar{\eta}_{n}=-\eta_{n}$.
 In terms
of the irreducible bath correlations (cumulants) $\tilde{C}_{\alpha_{N},\ldots,\alpha_{1}}^{\eta_{N},\ldots,\eta_{1}}$~\cite{GasbarriPRA2018},
the central system dynamics can be written as
\begin{equation}
\hat{\rho}^{{\rm S}}(t)=\mathcal{T}e^{\sum_{N=1}^{+\infty}\frac{2^N}{N!}\int_{0}^{t}dt_{N}\cdots dt_{1}\tilde{C}_{\alpha_{N},\ldots,\alpha_{1}}^{\eta_{N},\ldots,\eta_{1}}\mathcal{S}_{\alpha_{N}}^{\bar{\eta}_{N}}(t_{N})\cdots\mathcal{S}_{\alpha_{1}}^{\bar{\eta}_{1}}(t_{1})}\hat{\rho}^{{\rm S}}(0)\label{eq:irreducible}
\end{equation}
 Hereafter summation over the repeated indices $\eta_{n}$ and $\alpha_{n}$
is assumed.
The effects of a quantum bath
can be fully simulated by complex classical noises $b_{\alpha}(t)=b_{\alpha}^{+}(t)+ib_{\alpha}^{-}(t)$
that have the correlations $\left\langle b_{\alpha_{N}}^{\eta_{N}}(t_{N})\cdots b_{\alpha_{2}}^{\eta_{2}}(t_{2})b_{\alpha_{1}}^{\eta_{1}}(t_{1})\right\rangle =C_{\alpha_{N},\ldots,\alpha_{2},\alpha_{1}}^{\eta_{N},\ldots,\eta_{2},\eta_{1}}.$
The equivalence between a quantum bath and complex classical noises
in their effects on central system dynamics offers an interesting
venue for studying non-Hermitian quantum dynamics and thermodynamics
in complex plane~\cite{WeiPRL2012,PengPRL2015,WeiSciRep2014}.

\begin{figure}[b]
\includegraphics[width=0.9\columnwidth]{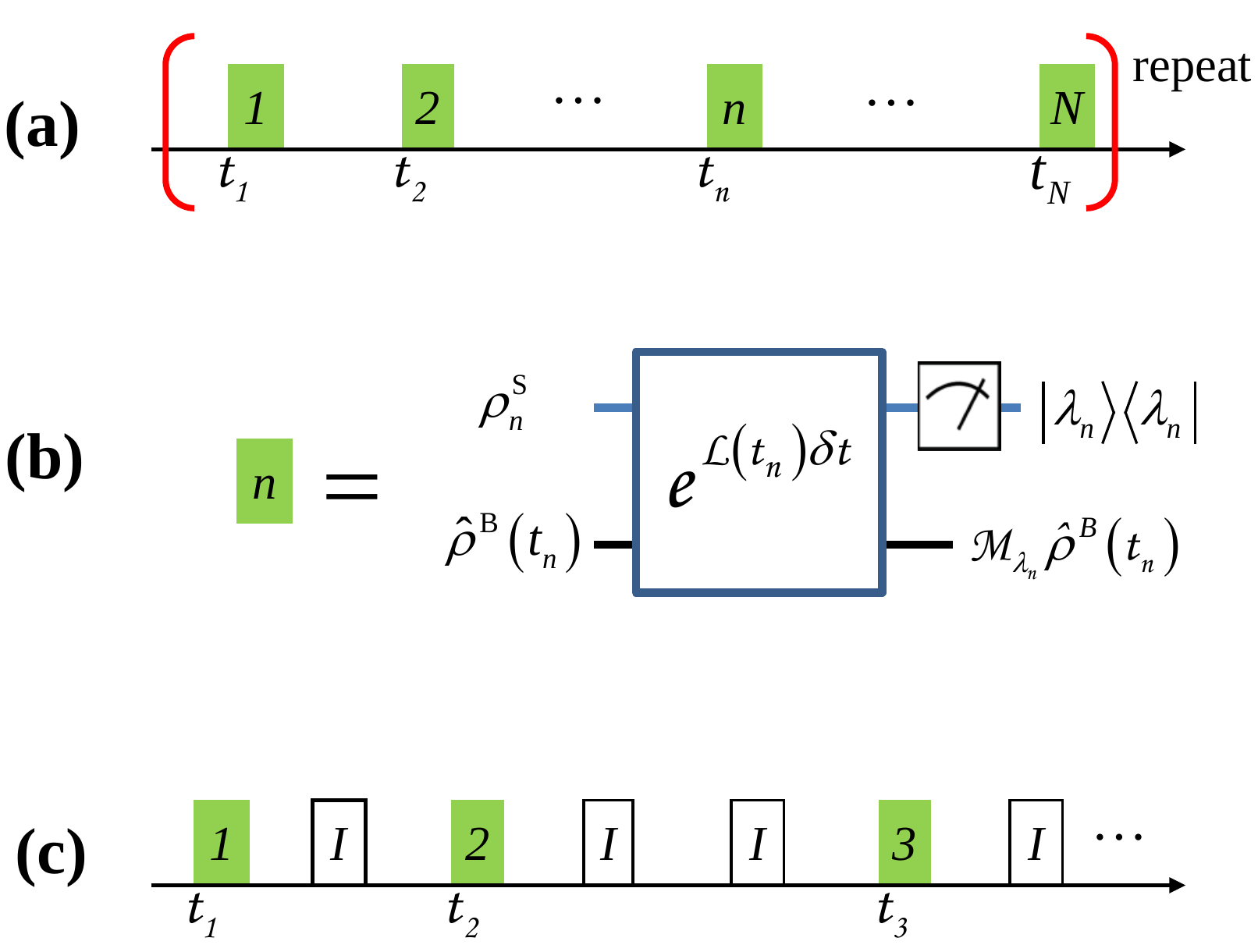}
\caption{Weak measurement for reconstruction of bath correlations. (a) A unit sequence (to be repeated many times) of $N$ weak measurements at different times for
reconstructing the bath correlations at the $N$-th order. (b) Realization of a weak measurement on the bath via the projective measurement of the central system.
(c) Reconstruction of the bath correlations by selecting a subset of outputs from a long measurement sequence, with the unused outputs in-between taken as ``idle'' (I). \label{fig_weak_measure}}
\end{figure}

\textit{Measurement of bath correlations} \textemdash{} We present
a protocol for measuring the bath correlations to an arbitrary order.
The scheme is based on weak measurement of the bath via
projective measurement of the central system (see Fig.~\ref{fig_weak_measure}).
To measure an $N$-th order correlation, a unit sequence of $N$ weak
measurements {[}Fig.~\ref{fig_weak_measure}(a){]}
is applied on the quantum bath. In each unit, the quantum bath is
prepared in the initial state $\rho^{{\rm B}}$ at $t=0$ and then
evolves under the bath Hamiltonian $H_{{\rm B}}$. At time $t_{n}$
(for $n=1,2,\ldots,N$), the central system is prepared in the state
$\rho_{n}^{{\rm S}}$, and then is coupled to the bath through the
interaction $V=\sum_{\alpha}S_{\alpha}B_{\alpha}$ for a small period
$\delta t$ of evolution. The state of the central system and the
bath, in the interaction picture, becomes $\hat{\rho}(t_{n}+\delta t)\approx e^{\mathcal{L}(t_{n})\delta t}\hat{\rho}^{{\rm B}}(t_{n})\otimes\rho_{n}^{{\rm S}}$,
where $\mathcal{L}(t)=2\sum_{\alpha}\left[\mathcal{S}_{\alpha}^{+}\mathcal{B}_{\alpha}^{-}(t)+\mathcal{S}_{\alpha}^{-}\mathcal{B}_{\alpha}^{+}(t)\right]$.
A quantity $\Lambda_{n}$ of the central system is measured at $t_{n}+\delta t$.
The output would be randomly an eigenvalue ${\lambda}_{n}$ of $\Lambda_{n}$
corresponding to the eigenstate $|{\lambda}_{n}\rangle$. The unit sequence of $N$ measurements
is repeated many times. The outputs of the $N$ measurements, averaged
over the repeated units, yield the measurement correlation $G^{(N)}=\left\langle {\lambda}_{N}\cdots {\lambda}_{2}{\lambda}_{1}\right\rangle $.

The projective measurement of the system operator $\Lambda_{n}$ constitutes
a weak measurement of the bath (due to the weak entanglement during
the interaction in the small period of time). See Fig.~\ref{fig_weak_measure}(b)
for illustration. The weak measurement is characterized by the Kraus
superoperator $\mathcal{M}_{{\lambda}_{n}}={\rm Tr}_{{\rm S}}\left[|{\lambda}_{n}\rangle\langle {\lambda}_{n}|e^{\mathcal{L}(t_{n})\delta t}\rho_{n}^{{\rm S}}\right]$
corresponding to the output ${\lambda}_{n}$. The probability of the output
${\lambda}_{n}$ is $p({\lambda}_{n})={\rm Tr}_{{\rm B}}\left[\mathcal{M}_{{\lambda}_{n}}\hat{\rho}^{{\rm B}}(t_{n})\right]$.
The bath state after the measurement is $\mathcal{M}_{{\lambda}_{n}}\hat{\rho}^{{\rm B}}(t_{n})/p({\lambda}_{n})$.
The joint probability of a sequence of $N$ outputs is $p({\lambda}_{N},\ldots,{\lambda}_{1})={\rm Tr}_{{\rm B}}\left[\mathcal{T}\mathcal{M}_{{\lambda}_{N}}\cdots\mathcal{M}_{{\lambda}_{1}}\rho^{{\rm B}}\right].$
The measurement correlation is $G^{(N)}=\sum_{\{{\lambda}_{n}\}}p({\lambda}_{N},\ldots,{\lambda}_{1}){\lambda}_{N}\cdots {\lambda}_{1}.$
For small $\delta t$, the evolution during the interaction $e^{\mathcal{L}(t_{n})\delta t}\approx1+\mathcal{L}(t_{n})\delta t$.
To pick up the signal  proportional to the noise fields ${B}_{\alpha}$ (hence proportional to the interaction $\mathcal{L}$),
we choose the initial state $\rho_{n}^{{\rm S}}$ and the observable
$\Lambda_{n}$ such that the background term ${\rm Tr}\left[\Lambda_{n}\rho_{n}^{{\rm S}}\right]=0$.
Thus, the measurement correlation up to the leading order of $\delta t$
is
\begin{equation}
G^{(N)}\approx \delta t^{N}\sum A_{\alpha_{N}}^{\bar{\eta}_{N}}(t_{N})\cdots A_{\alpha_{2}}^{\bar{\eta}_{2}}(t_{2})A_{\alpha_{1}}^{\bar{\eta}_{1}}(t_{1})C_{\alpha_{N},\ldots,\alpha_{2},\alpha_{1}}^{\eta_{N},\ldots,\eta_{2},\eta_{1}},\label{eq:corr of out}
\end{equation}
 where the coefficient
\begin{align}
A_{\alpha}^{\eta}(t_{n})=2{\rm Tr}_{{\rm S}}\left[\Lambda_{n}\mathcal{S}_{\alpha}^{\eta}\rho_{n}^{{\rm S}}\right].
\label{eq_coefficient}
\end{align}
Equation~(\ref{eq:corr of out})
defines a linear equation for the bath correlations of the $N$-th
order. Since the coefficients can be independently set by choosing the
system state $\rho_{n}^{{\rm S}}$ and the observable $\Lambda_{n}$,
a set of linearly independent equations can be established. By solving the set of linear
equations, the bath correlations can be reconstructed.

With the cumulant expansion in Eq.~(\ref{eq:irreducible}), only the
irreducible bath correlations are needed. Below we use the shorthand
notation $C(N,\ldots,1)\equiv C_{\alpha_{N},\ldots,\alpha_{1}}^{\eta_{N},\ldots,\eta_{1}}$.
The irreducible bath correlations (culmulants) $\tilde{C}(N,\ldots,2,1)$
can be recursively obtained by $\tilde{C}(1)=C(1)$, $\tilde{C}(2,1)={C}(2,1)-\tilde{C}(2)\tilde{C}(1)$,
$\tilde{C}(3,2,1)={C}(3,2,1)-\tilde{C}(3,2)\tilde{C}(1)-\tilde{C}(3,1)\tilde{C}(2)-\tilde{C}(2,1)\tilde{C}(3)-\tilde{C}(3)\tilde{C}(2)\tilde{C}(1)$,
and so on. The cumulant expansion can often be truncated at a rather
low order. In particular, in the case of Gaussian baths (such as a quardratic
boson bath~\cite{GasbarriPRA2018}), the truncation at the second order irreducible
correlations is exact. The truncation approximation would greatly
reduce the number of measurements required to reconstruct the bath
correlations.

We remark that both the ``classical'' and the ``quantum'' parts of the bath correlations can be extracted from the weak measurements.
The ``quantum'' correlations refer to the terms that contain at least one bath superoperator ${\mathcal B}^{\eta_n}_{\alpha_n}$ with $\eta_n=-$ (a
commutator) and the ``classical'' correlations contain only bath superoperators with $\eta_n=+$ (anti-commutators).
This classification of bath correlations is based on the observation that
the commutator ${\mathcal B}^{-}_{\alpha}$ would vanish if $B_{\alpha}$ is a classical noise field. As shown in Eq.~(\ref{eq_coefficient}),
to extract a quantum correlation, one just need to choose the central system state $\rho^{\rm S}_n$ and observable $\Lambda_n$ such that
$A_{\alpha}^{+}(t_{n})=2{\rm Tr}_{{\rm S}}\left[\Lambda_{n}\mathcal{S}_{\alpha}^{+}\rho_{n}^{{\rm S}}\right]\ne 0$.
It should be noted that for a bath at infinitely high temperature (such as a nuclear spin bath at room temperature~\cite{PfenderNC2019}), $\rho^{\rm B}\propto 1$ and $\mathcal{B}_{\alpha}^-\rho^{\rm B}=0$,
so the quantum correlations at the second order $C^{+,-}_{\alpha_2,\alpha_1}=0$. In this case, one needs to examine at least the third order to extract the quantum correlations in a quantum bath.
Such higher-order, ``quantum'' correlations are signatures of coherent clusters in baths~\cite{YangPRB2008a}. For example, these signatures can be employed for quantum sensing
of correlated nuclear spins in nuclear spin baths~\cite{YangPRB2008a,Ma2016PRApp}. The higher-order quantum correlations may also be used to study the quantum characteristics (such as the Leggett-Garg inequality~\cite{Leggett1985})
of quantum baths.

In practice, the protocol for reconstructing the bath correlations
can be simplified by exploiting the facts that the perturbation
of the weak measurement to the bath is small and the bath is usually
in a thermal equilibrium state. One can perform an indefinitely long sequence
of weak measurements on the bath at $t_k$ (with, e.g., $t_{k}=k\tau$
for $k=1,2,\ldots$). In each shot of measurement, the central system is prepared in state $\rho^{\rm S}_k$ at $t_k$, coupled to the bath through $V$ for time $\delta t$, and then measured on the
observable $\Lambda_k$ at $t_k+\delta t$. No preparation of the bath state is needed.  See Fig.~\ref{fig_weak_measure}(c) for illustration.  The measurement
correlations at a given order $N$ and for a given timing $(t_{1},t_{2},\ldots,t_{N})$
are obtained by selecting a subset of the measurements. The data from the other
measurements (taken as idle) are discarded (but would be used for constructing correlations
at other orders and/or for other timings).
For the measurement whose output  $\lambda_k$ is discarded, the evolution of the bath
averaged over all possible outputs, is $\sum_{{\lambda}_{k}}\mathcal{M}_{{\lambda}_{k}}\hat{\rho}^{{\rm B}}(t_{k})\equiv\mathcal{M}_{k}\hat{\rho}^{{\rm B}}(t_{k})$,
which amounts to measurement-induced decoherence.  If the measurement is weak ($\left|V\delta t\right|\ll1$), the measurement-induced decoherence is negligible, i.e.,
$\mathcal{M}_{k}\approx 1$ and
$\mathcal{M}_{k}\hat{\rho}^{{\rm B}}(t_{k})\approx \hat{\rho}^{{\rm B}}(t_{k})$.
Furthermore, if the bath is in the thermal
equilibrium state $\rho^{{\rm B}}\propto e^{-H_{{\rm B}}/(k_{B}T)}$
at temperature $T$, the bath Hamiltonian $H_{\rm B}$ induces no evolution on it.
Therefore, under the conditions that the bath is initially in a thermal equilibrium state and the measurement is sufficiently weak, the measurement correlations extracted from a subset of the measurements are the same,
in the leading order of $\delta t$, as those obtained without the idle measurements [that is, the same as those in Eq.~(\ref{eq:corr of out})].
In this simplified protocol,
the sequential weak measurements
can be carried out with a simple timing (e.g., equally spaced in time), there is no need to prepare the bath state in each unit sequence of measurement,
and the output data can be reused for constructing correlations at
different orders and for different timings~\cite{PfenderNC2019}.

\textit{Special case of central spin-$1/2$ \textemdash{}} As an example, we prepsent the explicit protocol
for reconstructing the correlations in a quantum bath of a central spin-1/2 (qubit).
The qubit-bath coupling can
be written as ${V}=\frac{1}{2}\sum_{\alpha=x,y,z}{\sigma}_{\alpha}{{}}{B}_{\alpha}$, where $\sigma_{\alpha}$ is the Pauli
matrix of the qubit along the $\alpha$-axis and
${{}}{B}_{\alpha}$ is the magnetic noise operator. Without loss of generality, we assume $t_N>t_{N-1}>\cdots > t_1$ in the correlation functions.

Let us consider the weak measurement of the bath at $t_1$ first.
The central spin is polarized to be along, e.g., the $x$-axis, described by the density operator
 ${{}}{\rho}^{\mathrm{S}}_1=(1+\sigma_x)/2$ at $t=t_1$. After the interaction with the bath through $V$ for time $\delta t$, a spin operator
$\Lambda_{1}$ is measured. To make the background term  $\mathrm{Tr}\left[\Lambda_1 {\rho}^{\mathrm{S}}_1\right]$ vanish, we choose the measurement axis to be along a
direction perpendicular to the initial polarization, e.g., $\Lambda_1=\sigma_y$.
With the definition in Eq.~(\ref{eq_coefficient}), the coefficient in Eq.~(\ref{eq:corr of out}) becomes
\begin{align}
A_{y}^{+}(t_1)=&{\rm Tr}\left[\sigma_y\left(\sigma_y\rho^{\rm S}_1+\rho^{\rm S}_1\sigma_y\right)\right]=1,
\nonumber \\
A_{z}^{-}(t_1)=&-i{\rm Tr}\left[\sigma_y\left(\sigma_z\rho^{\rm S}_1-\rho^{\rm S}_1\sigma_z\right)\right]=1,
\nonumber
\end{align}
and ${\rm else}=0$.
Therefore the measurement correlation becomes
\[
G^{(N)} = \delta t^{N} A_{{\alpha}_{N}}^{\bar{\eta}_{N}}(t_N)\cdots A_{\alpha_{2}}^{\bar{\eta}_{2}}(t_{2})\left(C_{\alpha_{N},\ldots, \alpha_{2},y}^{\eta_{N},\ldots,\eta_{2}, -}+C_{\alpha_{N},\ldots, \alpha_{2},z}^{\eta_{N},\ldots, \eta_{2} ,+}\right).
\]
Or if we choose $\rho^{\rm S}_1=(1-\sigma_x)/2$ (central spin initially polarized along the $-x$ direction) and $\Lambda_1=\sigma_y$, we have $A_{y}^{+}(t_1)=-A_{z}^{-}(t_1)=1$ and ${\rm else}=0$. The measurement correlation would be
\[
{\bar G}^{(N)} = \delta t^{N}  A_{{\alpha}_{N}}^{\bar{\eta}_{N}}(t_N)\cdots A_{\alpha_{2}}^{\bar{\eta}_{2}}(t_{2})\left(C_{\alpha_{N},\ldots, \alpha_{2},y}^{\eta_{N},\ldots,\eta_{2}, -}-C_{\alpha_{N},\ldots, \alpha_{2},z}^{\eta_{N},\ldots, \eta_{2} ,+}\right).
\]
The summation and difference of $G^{(N)}$ and ${\bar G}^{(N)}$ pick up the bath correlations $C_{\alpha_{N},\ldots, \alpha_{2},y}^{\eta_{N},\ldots,\eta_{2}, -}$ and $C_{\alpha_{N},\ldots, \alpha_{2},z}^{\eta_{N},\ldots, \eta_{2} ,+}$, respectively.
That is
\begin{subequations}
\begin{align}
G^{(N)}+{\bar G}^{(N)} = & {2\delta t^N}  A_{{\alpha}_{N}}^{\bar{\eta}_{N}}(t_N)\cdots A_{\alpha_{2}}^{\bar{\eta}_{2}}(t_{2}) C_{\alpha_{N},\ldots, \alpha_{2},y}^{\eta_{N},\ldots,\eta_{2}, -},
\label{G2Ca} \\
G^{(N)}-{\bar G}^{(N)}  = &  {2\delta t^N}  A_{{\alpha}_{N}}^{\bar{\eta}_{N}}(t_N)\cdots A_{\alpha_{2}}^{\bar{\eta}_{2}}(t_{2}) C_{\alpha_{N},\ldots, \alpha_{2},z}^{\eta_{N},\ldots,\eta_{2}, +}.
\label{G2Cb}
\end{align}
\label{eq:G2C}
\end{subequations}
 The procedure can be similarly applied to the measurements at other times. For the latest time $t_N$,
since the correlation function vanishes for $\eta_N=-$, only one set of ($\rho^{\rm S}_N,\Lambda_N$) is needed to pick up the correlation function $C_{\alpha_{N}\alpha_{N-1},\ldots,\alpha_1}^{+,\eta_{N-1},\ldots, \eta_1}$. Thus, using measurement correlation functions for $2^{N-1}$ configurations of central spin initialization and measurement directions $\left\{(\rho^{\rm S}_N,\Lambda_N)\right\}$, one can
determine $2^{N-1}$ bath correlation functions $C_{\alpha_{N}\alpha_{N-1}\ldots, \alpha_{1}}^{+,\eta_{N-1},\ldots, \eta_{1}}$ with $\alpha_n=y$ or $z$ corresponding to $\eta_n=-$ or $+$ for each $n$. The correlations of noise fields along other directions
can be similarly determined (e.g., correlations with $(\alpha_n,\eta_n)=(x/z,\pm)$ can be extracted from measurements with $\rho^{\rm S}_n=(1\pm \sigma_y)/2$ and $\Lambda_n=\sigma_z$).
For example, the third-order correlation (for $t_3>t_2>t_1$)
\begin{align}
 C^{+-+}_{x,y,z}=\left(G^{y,x,x}_{z,y,y}+G^{y,\bar{x},x}_{z,y,y}-G^{y,x,\bar{x}}_{z,y,y}-G^{y,\bar{x},\bar{x}}_{z,y,y}\right)/\left(4\delta t^3\right),
\nonumber
\end{align}
where $G^{y,\bar{x},x}_{z,y,y}$ denotes the measurement correlations for the central spin initialized along $x$ and measured along $y$ at $t_1$, initialized along $-x$ and measured along $y$ at $t_2$, and initialized along $y$ and measured along $z$ at $t_3$ (similarly for $G$ with other indices).

Among different types of decoherence, pure dephasing is often the most relevant to quantum information technology since it does not involve the slow energy dissipation process.
For pure dephasing, the qubit-bath coupling assumes the form $\hat{V}(t)=S_z\hat{B}_z(t)$. The qubit dynamics is determined by
the bath correlations as
\begin{align}
\hat{\rho}^{\mathrm{S}}(t)= \sum_{N=0}^{+\infty}\frac{2^{N}}{N!}\int_{0}^{t} dt_N\cdots dt_1 C_{z, \ldots, z}^{+\ldots+} {\mathcal S}_{z}^{-}\cdots {\mathcal S}_{z}^{-}\hat{\rho}^{\mathrm{S}}(0).
\nonumber
\end{align}
Here we have used the fact that  $C_{z,z,\ldots,z}^{-\eta_{N-1}\ldots \eta_{1}}=0$ and
$\mathcal{S}_z^{+}\mathcal{S}_z^{-}=0$.
In the case of pure dephasing, the effects of quantum bath is fully determined by the correlation $C_{z, \ldots, z}^{+\ldots+}$, which
is directly related to the weak measurement correlations through, e.g.,
$$
C^{+++}_{z,z,z}=G^{x,x,x}_{y,y,y}/\delta t^3.
$$
Here we have used the fact that in the pure dephasing case, $C^{+-+}_{z,y,z}=C^{++-}_{z,z,y}=C^{+--}_{z,y,y}=0$ (for $B_y=0$) and therefore $G^{x,x,x}_{y,y,y}=-G^{x,\bar{x},x}_{y,y,y}=
-G^{x,x,\bar{x}}_{y,y,y}=G^{x,\bar{x},\bar{x}}_{y,y,y}$ [according to Eq.~(\ref{eq:G2C})].
It should be noted that even though the pure dephasing is determined only by the ``classical'' bath correlations $C_{z, \ldots, z}^{+\ldots+}$, the ``quantum'' correlations (those that contain at least one commutator) can still be measured by weak measurements. For example,
$$
C^{+-+}_{z,z,z}=G^{x,x,x}_{y,z,y}/\delta t^3.
$$

\textit{Conclusion} \textemdash{} We propose a general scheme for complete
characterization of arbitrary order correlations in a quantum bath,
based on weak measurement of the bath realized by projective measurement of
a central system embedded in the bath. From the weak measurement
correlations at the $N$-th order, one can reconstruct the $N$-th order bath correlations. The weak measurement has the advantage
of negligible disturbance (i.e., measurement-induced decoherence)
to the bath - this advantage allows the measurement data be collected
at a simple timing and the correlations be extracted by selecting
certain subsets of the data, which greatly reduces the time consumption
for reconstructing the correlation functions~\cite{PfenderNC2019}.
Once the bath correlations are characterized, they can be used for
optimizing quantum controls under all circumstances~\cite{KofmanPRL2001,KofmanPRL2004,CywinskiPRB2008,GraceJPB2007,GordonPRL2008,WestPRL2010,LiuNatCommun2013}.
Characterizing arbitrary-order correlations in quantum baths may provide an approach to studying the quantum charactieristics (such as the Leggett-Garg inequality~\cite{Leggett1985}) of many-body environments and enable quantum sensing of nuclear spin
clusters of different types of correlations~\cite{YangPRB2008a,Ma2016PRApp}.
We expect that
the experimental demonstration of the protocol is feasible in solid
spin systems such as nitrogen-vacancy center spins~\cite{DohertyPhysRep2013},
donor spins in silicon~\cite{GeorgePRL2010} and quantum dots~\cite{HansonRMP2007}.

\acknowledgements{This work was supported by Hong Kong RGC-NSFC Joint Scheme - Project N\_CUHK403/16, National Key Research and Development Program of China (No. 2018YFA0306600), National Natural Science Foundation of China (Nos. 11425523, 11661161018), and the European Union funding via SMeL and ASTRIQS as well as the Volkswagen Foundation.}

\end{document}